\documentclass[aps,prl,reprint,groupedaddress,showpacs]{revtex4-1}

\usepackage{graphicx}
\usepackage{amsmath}

\newcommand{\fbare}{f_\mathrm{bare}}
\newcommand{\fdressed}{f_\mathrm{dressed}}

\newcommand{\Tone}{T_1}
\newcommand{\Ttwo}{T_2^\mathrm{echo}}
\newcommand{\Gammatwo}{\Gamma_2}

\newcommand{\tRamsey}{t_\mathrm{R}}
\newcommand{\tbuffer}{t_\mathrm{buffer}}
\newcommand{\trelax}{t_\mathrm{relax}}
\newcommand{\tgate}{t_\mathrm{gate}}
\newcommand{\tMone}{t_{M1}}
\newcommand{\tMtwo}{t_{M2}}
\newcommand{\Tcav}{T_\mathrm{cav}}
\newcommand{\Pnorm}{P_\mathrm{norm}}

\newcommand{\mK}{\mathrm{mK}}
\newcommand{\K}{\mathrm{K}}
\newcommand{\kHz}{\mathrm{kHz}}
\newcommand{\MHz}{\mathrm{MHz}}
\newcommand{\GHz}{\mathrm{GHz}}

\newcommand{\us}{\mu\mathrm{s}}
\newcommand{\ns}{\mathrm{ns}}

\newcommand{\Dw}{\Delta}
\newcommand{\no}{n_0}
\newcommand{\nbar}{n}

\begin{document}

\title{Rapid Driven Reset of a Qubit Readout Resonator}

\author{D. T. McClure}
\author{Hanhee Paik}
\author{L. S. Bishop}
\author{M. Steffen}
\author{Jerry M. Chow}
\author{Jay M. Gambetta}

\affiliation{IBM T.J. Watson Research Center, Yorktown Heights, NY 10598, USA}

\date{\today}

\begin{abstract}
Using a circuit QED device, we demonstrate a simple qubit measurement
pulse shape that yields fast ring-up and ring-down of the readout resonator
regardless of the qubit state. The pulse differs from a square pulse only by the
inclusion of additional constant-amplitude segments designed to effect a
rapid transition from one steady-state population to another.
Using a Ramsey experiment performed shortly after the measurement pulse to
quantify the residual population, we find that compared to a square pulse
followed by a delay, this pulse shape reduces the timescale for cavity ring-down
by more than twice the cavity time constant. At low drive powers, this
performance is achieved using pulse parameters calculated from a
linear cavity model; at higher powers, empirical optimization of the pulse
parameters leads to similar performance.
\end{abstract}

\pacs{}

\maketitle

Over the last decade, circuit quantum electrodynamics~\cite{blais_cavity_2004}
(cQED) has become a leading architecture for constructing scalable networks of
solid-state qubits, finding application in the context of not only
superconducting qubits~\cite{kelly_state_2015,corcoles_demonstration_2015} but
also spin qubits~\cite{petersson_circuit_2012} and potentially other
systems~\cite{muller_detection_2013}.
In this paradigm, each qubit is coupled to a resonator in which it induces a
state-dependent frequency shift, allowing the qubit state to be interrogated
using a pulsed tone near the resonator frequency. A great deal of research has
focused on optimizing the speed and fidelity of such measurements.
Most significantly, the ongoing development of quantum-limited
amplifiers~\cite{bergeal_phase-preserving_2010,vijay_observation_2011,ho_eom_wideband_2012,
hover_superconducting_2012} has improved achievable signal-to-noise ratios
enormously. The introduction of Purcell
filters~\cite{reed_fast_2010,jeffrey_fast_2014} has enabled the use of
resonators with fast time constants, whose high bandwidth would otherwise
provide a pathway for qubit relaxation via spontaneous emission (Purcell
effect~\cite{purcell_1946}). Some work~\cite{jeffrey_fast_2014,liu_high_2014}
has also explored the use of pulse shapes with an initial overshoot in order
to populate the readout resonator more quickly than the standard square pulse.

Although these improvements have made fast, high-fidelity qubit readout in cQED
systems routine, relatively little attention has been devoted to the problem of
returning the readout resonator to its ground state immediately after the
measurement pulse. If the pulse is simply turned off, residual photons gradually
exiting the resonator will continue to measure the
qubit~\cite{gambetta_qubit-photon_2006}, preventing high-fidelity operations for
a period of several time constants. Even for a resonator with a fast time
constant, this delay is typically longer than the time needed for qubit control
and measurement. A technique for reducing the residual population on a timescale
faster than the resonator's free decay is therefore desirable in any
algorithm in which qubits need to be re-used shortly after measurement, e.g.
error correction with surface~\cite{kitaev_fault_1997,bravyi_quantum_1998},
C4~\cite{aliferis_fibonacci_2009} or Bacon-Shor~\cite{aliferis_subsystem_2007}
codes. A major impediment has been the fact any such scheme needs to work in the
absence of prior knowledge as to which of the two possible state-dependent
resonator frequencies will be realized.

\begin{figure}[b]
\includegraphics[width=3.4in]{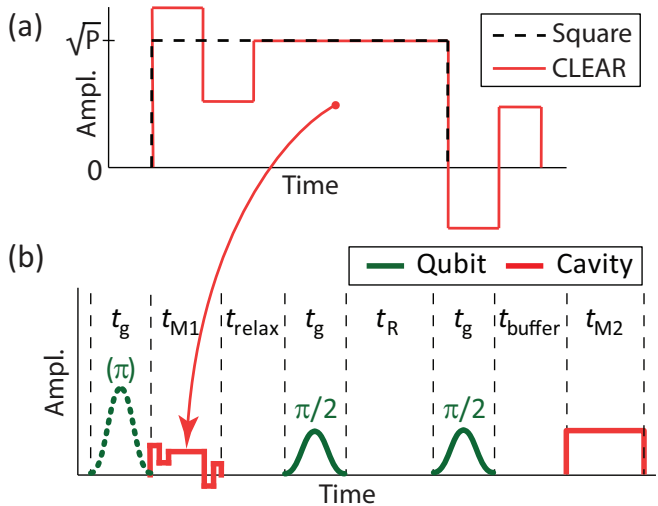}
\caption{\label{CLEARTestSequence} (color) (a) Schematic shape of the
piecewise-constant CLEAR pulse (solid red line) with a square pulse
(dashed black line) for reference. (b) Pulse sequence used to extract the
residual cavity population after a measurement pulse.
The qubit is prepared in either the ground or excited state and
then a measurement pulse (either a square pulse or a CLEAR pulse) of length
$\tMone$ is applied. After the measurement pulse, an adjustable
delay $\trelax$ precedes a pair of $X_{90}$ pulses separated by
$\tRamsey$ comprising a Ramsey experiment. The Ramsey experiment is followed by
another brief delay $\tbuffer$ and a square measurement pulse of length
$\tMtwo = 10~\us$.}
\end{figure}

Here we present the first experimental demonstration of driven
state-independent reset of a readout resonator, using a specially designed yet
simple pulse shape that we term the Cavity Level Excitation and Reset
(CLEAR) pulse [Fig.~\ref{CLEARTestSequence}(a)]. The pulse uses short segments to ``kick''
the resonator rapidly from one steady-state population to another: at the
beginning of the pulse, two such segments drive the population from zero to the desired steady-state value, and
at the end, two more drive it back to zero. In this work, we focus on
quantifying the effectiveness of the depopulating segments. Using a Ramsey experiment to
extract the number of residual photons in the cavity after the pulse, we
compare the performance to that of a standard square pulse. We find that for
pulse powers where the cavity response remains linear, the theoretically
derived CLEAR pulse shape depopulates the cavity to a negligible level in a time
more than two cavity time constants faster than that needed after the square pulse. At higher powers,
optimizing the pulse shape empirically using an iterative algorithm leads to
equally good performance.

The experimental device is a fixed-frequency transmon qubit mounted in a 3D
aluminum cavity~\cite{paik_observation_2011} attached to the mixing chamber of a
dilution refrigerator at an indicated base temperature of $10~\mK$. Qubit and
measurement drive tones are generated using Agilent E8267D function generators
and modulated using Tektronix AWG7000 series arbitrary waveform generators at a
2 GS/s sample rate. Qubit pulses are $4\sigma$ Gaussians with
DRAG~\cite{motzoi_simple_2009} correction. The cavity is measured in
transmission, and the transmitted signal is fed to a HEMT amplifier (Low Noise
Factory LNF-LNC6\_20A) at $4~\K$ using a superconducting NbTi/NbTi semi-rigid coaxial cable. After additional
amplification at room temperature, the signal is mixed down to $16~\MHz$ and
digitally demodulated. The cavity is measured to have bare frequency
$\fbare=10.7457~\GHz$, dressed frequency $\fdressed = 10.7594~\GHz$, and
linewidth $\kappa/2\pi=1.1~\MHz$ (corresponding to a time constant $\Tcav=1/\kappa=0.14~\us$). The qubit has
frequency $f_{01}=4.83315~\GHz$, anharmonicity $\delta/2\pi=-155~\MHz$,
average $\Tone\approx50~\us$, and average $\Ttwo\approx60~\us$. Preparing the
qubit in the excited state shifts $\fdressed$ by the cavity pull $2\chi/2\pi=-2.6~\MHz$; the measurement tone is applied at the
midpoint of the two frequencies, $\fdressed-\chi/2\pi$.

The residual population after a measurement pulse can be quantified in terms of
the mean cavity photon number $\nbar$ at some time after the end of the pulse.
We use the sequence illustrated in Fig.~\ref{CLEARTestSequence}(b) to extract
$\nbar$ following an initial measurement pulse denoted M1. The measurement pulse
is followed by a quick Ramsey experiment ($\tgate = 8~\ns$, $\tRamsey = 0$ to $600~\ns$) to
probe the ac Stark shift and dephasing from any residual
photons~\cite{schuster_ac_2005,gambetta_qubit-photon_2006}.
The time $\trelax$ between the end of M1 and the start of the Ramsey experiment
can be varied to measure $\nbar$ as a function of time after the end of M1.
The time $\tbuffer$ between the Ramsey experiment and the second measurement
pulse (M2) is set to $400~\ns$ to ensure that even when $\tRamsey$ and $\trelax$
are both short, M2 is not corrupted by lingering photons from M1.

\begin{figure}[b] \includegraphics[width=3.4in]{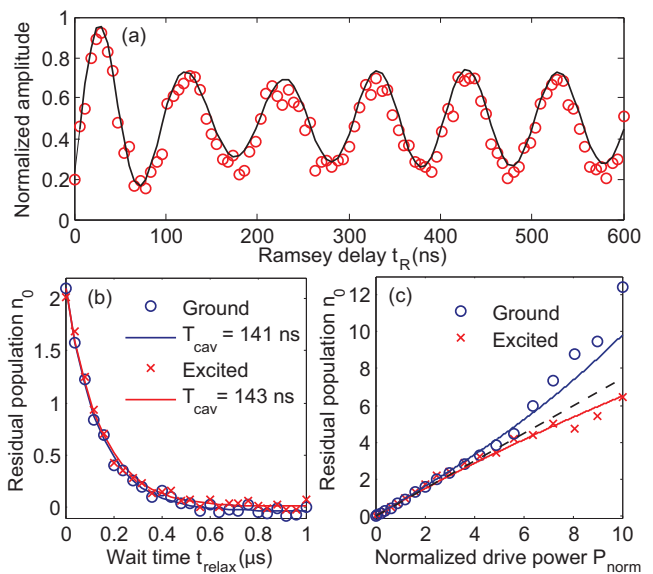}
\caption{\label{RamseyFits} (color) (a) Sample Ramsey experiment and fit. The
black curve is a fit of Eq.~\ref{RamseyEqn} to the data (red circles), yielding
initial cavity population $\no\approx0.9$. (b) Markers indicate $\no$ versus
wait time $\trelax$ after a square measurement pulse with drive power $\Pnorm=2$. Here and throughout, blue and red denote
experiments in which the qubit was prepared in the ground and excited states, respectively. Solid
blue and red curves are exponential fits to the respective data sets. (c) $\no$
versus drive power measured at $\trelax=40~\ns$ after a square
measurement pulse. The dashed line indicates the prediction for a linear
cavity, accounting for $\trelax$. The solid curves account for the
cavity's self-Kerr nonlinearity calculated from the measured qubit and cavity
parameters.}
\end{figure}

A typical Ramsey trace is shown in Fig.~\ref{RamseyFits}(a). The non-monotonic
modulation in both amplitude and frequency arises from the fact that the cavity
population evolves during the Ramsey delay, leading to recurrences. We derive
the expected form of the Ramsey trace during this transient response using the
positive-$P$-function method as in Gambetta et
al.~\cite{gambetta_qubit-photon_2006}, where it was applied to the steady-state
problem. For a Ramsey detuning $\Dw$ (here $10~\MHz$), decoherence rate
$\Gammatwo$, and initial phase $\phi_0$, the resulting functional form is
\begin{equation}
\label{RamseyEqn}
S(\tRamsey) = \frac{1}{2}[1 - \mathrm{Im}(\exp(-(\Gammatwo+\Dw i)\tRamsey +
(\phi_0-2\no\chi\tau) i))],
\end{equation}
where $\tau = (1-e^{-(\kappa+2\chi i)\tRamsey})/(\kappa+2\chi i)$ and
$\no$ is the value of $\nbar$ at the beginning of the Ramsey experiment. Using $\kappa$ and $\chi$ obtained from frequency-domain measurements
and taking $\Gammatwo=1/\Ttwo$, the only free parameters are $\no$ and
$\phi_0$. As illustrated in the figure, this function yields a good fit to the
data, allowing reliable determination of $\no$. In the rest of this work, we use
the extracted $\no$ to quantify the residual population as a function of pulse
shape, drive power, and wait time. We note that $\no$ does not include
the background thermal population of the cavity, which is accounted for by the
$\Gammatwo$ term and is calculated from $\Ttwo$ to be $\sim0.02$ on
average (assuming thermal photons are the dominant source of steady-state
dephasing~\cite{rigetti_superconducting_2012,sears_photon_2012}).

The Ramsey fit method of obtaining $\no$ was validated by using it to measure
$\no$ as a function of both wait time $\trelax$ and pulse power. For simplicity,
these tests were performed using square pulses for both M1 and M2, varying only
the power of M1, denoted $P$ as illustrated in Fig.~\ref{CLEARTestSequence}(a).
For convenience we define the normalized drive power $\Pnorm=
P/P_{1\mathrm{ph}}$, where $P_{1\mathrm{ph}}$ is the steady-state drive power
that yields $\nbar=1$, as inferred from a standard Ramsey experiment measuring
the Stark shift $\Delta\omega=2\chi\nbar$ \cite{gambetta_qubit-photon_2006}
induced by a CW tone. Figure~\ref{RamseyFits}(b) shows $\no$ extracted from
Ramsey fits as a function of $\trelax$. Regardless of the prepared qubit state,
the decay is exponential, as expected for free decay, and the time constant
$\Tcav$ extracted from the best-fit curve is consistent with the value of
$\kappa$ obtained from frequency-domain measurements.
Figure~\ref{RamseyFits}(c) shows $\no$ as a function of $\Pnorm$ at
$\trelax=40~\ns$; similar behavior was observed at all values of $\trelax$ for
which non-negligble $\no$ were measured. The dashed line indicates the expected
behavior assuming a linear cavity: $\no=e^{-\kappa\trelax}\Pnorm$.
The data exhibit a transition from a linear response at low powers to a super-
\mbox{(sub-)} linear response at high powers when the qubit is prepared in the
ground (excited) state. This behavior is consistent with the cavity's expected
self-Kerr nonlinearity~\cite{boissoneault_improved_2010}, which shifts the
cavity frequency by a negative amount $K$ per cavity photon, pushing it closer
to (farther from) the measurement frequency when the qubit is in the ground
(excited) state. Approximating $K$ in the small-$\delta$ limit as
$K=2g^4\delta(3\omega_q^4+2\omega_q^2\omega_r^2+3\omega_r^4)/(\omega_q^2-\omega_r^2)^4$,
where $\omega_q=2\pi f_{01}$, $\omega_r=2\pi\fdressed$, and $g$ is
calculated as in Ref.~\cite{rigetti_superconducting_2012}, we obtain $K\approx-14~\kHz$. The
solid blue (red) curve indicates the expected cavity response with the qubit
in the ground (excited) state, accounting for the calculated non-linearity.

Having thus validated our method of quantifying residual photons, we then
switched to a CLEAR pulse shape for M1; for consistency, we continued to use a
square pulse for M2. For an ideal single qubit-cavity system, the optimal CLEAR
pulse envelope [Fig.~\ref{CLEARTestSequence}(a)] consists of five
piecewise-constant segments: two ring-up segments, one steady-state segment, and
two ring-down segments. For the pulse to behave as intended, its bandwidth needs
only to be much greater than that of the cavity, a condition readily achieved in
our setup. Setting the carrier frequency to $\fdressed-\chi/2\pi$, as done here,
is not required but maximizes both SNR and simplicity: in this case, a given measurement pulse
yields the same steady-state $\nbar$ regardless of qubit state as long as the
cavity remains in the linear-response regime. The lengths of the ring-up and
ring-down segments were all initially fixed at $150~\ns$ (approximately
$\Tcav$), and their amplitudes relative to that of the steady-state segment
were calculated by solving a driven damped harmonic oscillator model to find the
pulse shape that populates and depopulates the cavity in the shortest amount of
time regardless of the qubit state.

\begin{figure}[b] \includegraphics[width=3.4in]{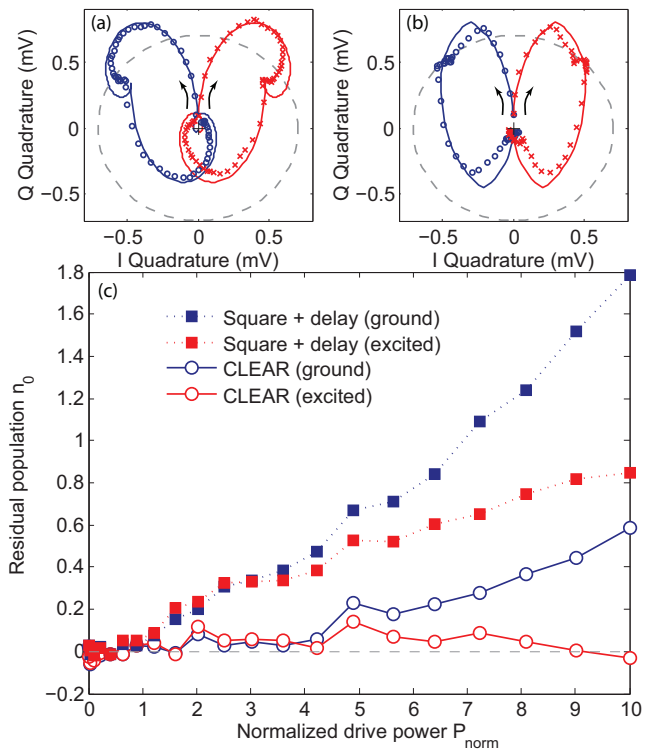}
 \caption{\label{CLEARvsSquare} (color) (a,b) IQ-plane cavity trajectories in
 response to a square pulse and CLEAR pulse, respectively. In each
 plot, experimental results (markers) are superimposed on theoretical
 calculations (solid curves), the dashed circle indicates the target population
 $\nbar=3.6$, the black cross indicates the origin, and the black arrows indicate the directions of the trajectories. (c) Residual cavity population versus pulse power for both
 square and CLEAR pulse shapes. For the CLEAR pulse, the Ramsey experiment begins immediately at the end of the pulse, while
 for the square pulse, a delay of approximately $300~\ns$ is inserted to match
 the total length of the CLEAR pulse's two ring-down segments.}
\end{figure}

The cavity IQ plane trajectories produced by square and CLEAR pulses with the
same steady-state amplitude ($\Pnorm = 3.6$) are shown by markers in
Figs.~\ref{CLEARvsSquare}(a,b), respectively. For all trajectories, the time
step between markers is $24~\ns$. Solid lines in these plots indicate the
theoretically calculated response of the cavity to each pulse, multiplied by an
overall amplitude factor to match the data and adjusted to reflect the
independently observed $20\%$ thermal population of the qubit excited state
(which was reduced by an order of magnitude on later cooldowns of
this device with additional input line attenuation). We see that the experimental
cavity responses track the theoretically calculated ones very well, and that
compared to the square pulse, the CLEAR pulse yields more compact trajectories that reach
near-steady-state populations (at both $\nbar\approx3.6$ and $\nbar\approx0$) in less time.

\begin{figure}[h!] \includegraphics[width=3.4in]{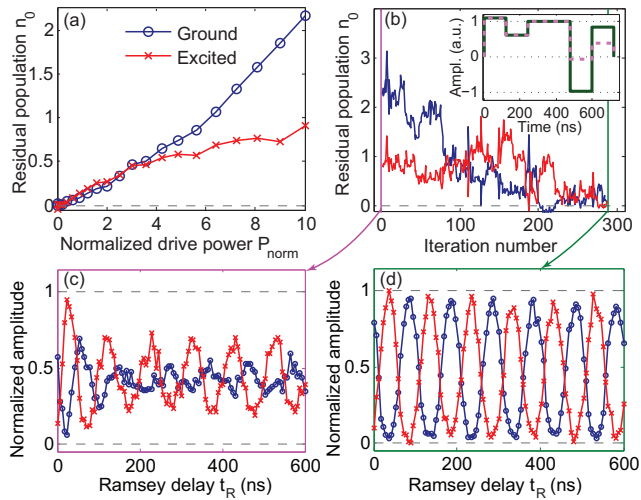}
\caption{\label{HighPowerCLEAR} (color) (a) $\no$ versus drive power for
the shortened CLEAR pulse ($120~\ns$ ring-down segments), for $\trelax=0$. (b)
Evolution of $\no$ at each step of an empirical optimization algorithm for the shortened CLEAR pulse with
$\Pnorm=10$. Inset: shortened CLEAR pulse shape before optimization (magenta)
and after (green). (c) Ramsey traces obtained using initial shortened CLEAR
pulse, yielding $\no\approx 2.2$ for ground and $\no\approx 0.91$ for
excited. (d) Ramsey traces obtained using final shortened CLEAR pulse,
yielding $\no < 0.1$ for both qubit states.}
\end{figure}

We quantitatively compare the performance of the CLEAR pulse to that of a square
pulse using the Ramsey fit method. To provide a fair comparison, a
zero-amplitude segment is appended to the square pulse to allow undriven decay
during a time equivalent to the total length of the CLEAR pulse's two ring-down
segments. The results are shown in Fig.~\ref{CLEARvsSquare}(c). At all
measurement powers, the CLEAR pulse significantly outperforms the square pulse;
moreover, for drive powers that keep the cavity in the linear regime (evidenced
by $\no$ being independent of the prepared qubit state), the residual population
immediately after the CLEAR pulse is negligible. As seen in
Fig.~\ref{RamseyFits}(b), for drive powers in this range, negligible $\no$ is
not obtained until $\sim600~\ns$ after a square pulse; allowing for the
$300~\ns$ taken by the ring-down segments of the CLEAR pulse, we find a net
speedup of $\sim300~\ns$, or approximately $2\Tcav$. At higher powers, it
appears that the cavity nonlinearity, not taken into account in calculating the optimal CLEAR pulse parameters, prevents perfect ring-down. We also find non-idealities when the lengths of the ring-down segments are reduced in an effort to shorten the ring-down time: for $120~\ns$ ring-down segments (a $20\%$ reduction), we find measurable $\no$ even in the linear regime, as illustrated in
Fig.~\ref{HighPowerCLEAR}(a). Further reductions in the segment lengths increase
the performance degradation.

To improve performance of the CLEAR pulse both at high powers and with shortened
ringdown segments, we use an empirical technique to optimize the pulse
parameters. As an example, we take as a starting point the pulse with $120~\ns$
ring-down segments and a steady-state drive power of $\Pnorm=10$ (which yielded
$\no\approx 2.16$ for the ground state and $\no\approx 0.91$ for the excited
state), and run an iterative optimization algorithm that attempts to minimize
$\no$ by adjusting the amplitudes of the ring-down segments. We keep
$\trelax=0$ throughout. The evolution of $\no$ with each iteration is shown in
Fig.~\ref{HighPowerCLEAR}(b): in fewer than 300 iterations, the pulse is
optimized to yield $\no<0.1$ regardless of initial qubit state.
Ramsey experiments before and after running this optimization are shown in
Fig.~\ref{HighPowerCLEAR}(c,d), revealing that coherence is preserved with the
optimized pulse shape regardless of initial qubit state. As seen in the inset of
Fig.~\ref{HighPowerCLEAR}(b), the optimization process significantly increases
the amplitudes of both ring-down segments. Extending our theoretical
calculations to the non-linear regime may shed light on this result and
potentially eliminate the need for empirical tune-up in this regime.

In summary, using a single-qubit cQED system, a qubit-state-independent
reduction in the time needed to reach a steady-state resonator population both
during and after a qubit measurement pulse was achieved by including extra
constant-amplitude segments in the pulse. For low-power drives, near-perfect
ring-down (quantified using Ramsey experiments) is achieved using segment
amplitudes calculated from system parameters. At higher drive amplitudes,
similar performance is obtained following empirical optimization of the pulse
shape. Though this demonstration used a 3D transmon, the same technique should
be applicable to any cQED system; it may also be combined with machine-learning
based analysis~\cite{magesan_machine_2014} and Purcell filters to further reduce
the measurement cycle time. Future areas of interest may include numerical
calculation of the optimal CLEAR parameters in the non-linear regime, extension
of this technique to resonators coupled to multiple qubits, and investigation as
a possible method for implementing the resonator-induced phase (RIP)
gate~\cite{cross_optimized_2014} non-adiabatically.

We thank M. B. Rothwell and G. A. Keefe for device fabrication, J. R. Rozen and
J. Rohrs for technical assistance, and B. Abdo for useful discussions. We
acknowledge support from IARPA under Contract No. W911NF-10-1-0324. All
statements of fact, opinion or conclusions contained herein are those of the
authors and should not be construed as representing the official views or
policies of the U.S. Government.


\bibliography{CLEARRefs}

\end{document}